\documentclass[a4paper,11pt]{article}
\pdfoutput=1
\usepackage{jheppub}

\usepackage{amssymb,amsmath}
\usepackage[normalem]{ulem}
\usepackage[utf8x]{inputenc}
\usepackage{slashed}
\usepackage{graphicx}
\usepackage{here}
\usepackage{color}
\usepackage{csquotes} 
\usepackage{comment}
\usepackage{mathrsfs}
\usepackage{float}
\usepackage{ascmac}
\usepackage{multirow}
\usepackage{longtable}

\usepackage[italicdiff]{physics}

\usepackage[hang,small,bf]{caption}
\usepackage[subrefformat=parens]{subcaption}
\usepackage{hyperref}

\usepackage{ulem}
\usepackage{xcolor}

\newcounter{c1}
\newcounter{c}
\newcounter{c2}
\newcounter{c3}
\newcounter{c4}
\DeclareRobustCommand{\erase}{\bgroup\markoverwith{\textcolor{blue}{\rule[.5ex]{2pt}{0.4pt}}}\ULon}

\makeatletter
\newcommand*\rel@kern[1]{\kern#1\dimexpr\macc@kerna}
\newcommand*\widebar[1]{%
  \begingroup
  \def\mathaccent##1##2{%
    \rel@kern{0.8}%
    \overline{\rel@kern{-0.8}\macc@nucleus\rel@kern{0.2}}%
    \rel@kern{-0.2}%
  }%
  \macc@depth\@ne
  \let\math@bgroup\@empty \let\math@egroup\macc@set@skewchar
  \mathsurround\z@ \frozen@everymath{\mathgroup\macc@group\relax}%
  \macc@set@skewchar\relax
  \let\mathaccentV\macc@nested@a
  \macc@nested@a\relax111{#1}%
  \endgroup
}
\makeatother

\numberwithin{equation}{section}

\preprint{
\begin{minipage}{5cm}
\small
\flushright
EPHOU-25-004\\KYUSHU-HET-318
\end{minipage}}

\title{Coupling Selection Rules in Heterotic Calabi-Yau Compactifications}

\author{Jun Dong$^{1}$,} 
\author{Tatsuo Kobayashi$^{1}$,} 
\author{Ryusei Nishida$^{1}$,} 
\author{Satsuki Nishimura$^{2}$, \\ and } 
\author{Hajime Otsuka$^{2}$} 
\affiliation{
$^1$Department of Physics, Hokkaido University, Sapporo 060-0810, Japan\\
$^2$Department of Physics, Kyushu University, 744 Motooka, Nishi-ku, Fukuoka 819-0395, Japan
}
\emailAdd{j-dong@particle.sci.hokudai.ac.jp}
\emailAdd{kobayashi@particle.sci.hokudai.ac.jp}
\emailAdd{r-nishida@particle.sci.hokudai.ac.jp}
\emailAdd{nishimura.satsuki@phys.kyushu-u.ac.jp}
\emailAdd{otsuka.hajime@phys.kyushu-u.ac.jp}

\abstract{
We study coupling selection rules of chiral matter fields in heterotic string theory with standard embedding.
These selection rules are determined by topological properties of Calabi-Yau threefolds.
We classify coupling selection rules on complete intersection Calabi-Yau threefolds for $h^{1,1}\leq 5$.
It is found that all of these selection rules for $h^{1,1}\leq 5$ are understood by combinations of only five types of fusion rules.
}
\makeatletter
\gdef\@fpheader{}
\makeatother

\begin{document}

\maketitle

\section{Introduction}
\label{sec:intro}

Symmetries are quite important in particle physics and string theory.
They determine possible couplings among particles and forbid certain couplings, that is, coupling selection rules.
One often discusses the coupling selection rules due to group theory, but there are many examples exhibiting selection rules without group action. 
For instance, when the theory has non-invertible symmetries, they lead to new selection rules, requiring more than a group-like structure (see Refs.~\cite{Schafer-Nameki:2023jdn,Shao:2023gho} for reviews on non-invertible symmetries).

It was indeed known that string compactifications lead to non-trivial selection rules.
For example, in heterotic string theory on toroidal orbifolds, some selection rules can be understood by group theory such as gauge invariance, Lorentz invariance of compact space, and R-symmetry. 
Discrete flavor symmetries such as $D_4$ and $\Delta (54)$ also appear \cite{Dijkgraaf:1987vp,Kobayashi:2004ya,Kobayashi:2006wq,Beye:2014nxa}. 
Recently, the discrete flavor symmetry $D_4$ was discussed from the viewpoint of non-invertible symmetry \cite{Thorngren:2021yso,Kaidi:2024wio,Heckman:2024obe}. 
It turned out that the coupling selection rules can not be simply understood by group theory \cite{Font:1988nc,Kobayashi:2011cw}.

Toroidal compactifications with magnetic fluxes also lead to similar discrete flavor symmetries such as $D_4$ and $\Delta (27) $ \cite{Abe:2009vi,Berasaluce-Gonzalez:2012abm,Marchesano:2013ega}.
Recently, it is found that non-invertible (flavor) symmetries appear from $\mathbb Z_2$ orbifolding of magnetized toroidal compactifications \cite{Kobayashi:2024yqq,Funakoshi:2024uvy}.
That includes a gauging of the outer automorphism of $\mathbb Z_M$ symmetries, i.e., $D_M\cong \mathbb{Z}_M\rtimes \mathbb{Z}_2$ with gauged $\mathbb{Z}_2$.\footnote{For gauging of the outer automorphism of a group, see, Refs. \cite{Heidenreich:2021xpr,Bhardwaj:2022yxj}.} 
These non-invertible flavor symmetries were applied to flavor physics in order to derive interesting flavor structure \cite{Kobayashi:2024cvp,Kobayashi:2025znw}. 
Such a $\mathbb Z_2$ gauging of $\mathbb Z_M$ can lead to non-trivial Yukawa textures, which can not be realized by group selection rules without symmetry breaking.
For example, the Yukawa texture of the nearest neighbor interaction pattern and other interesting textures can be obtained. It includes the texture addressing the strong CP problem without axion.

Coupling selection rules in heterotic Calabi-Yau (CY) compactifications with standard embedding are determined by topological data of CY such as intersection numbers \cite{Candelas:1985en}.
In general, such coupling selection rules are different from 
coupling selection rules due to group theory. 
For instance, non-invertible global symmetries are discussed in two-dimensional superconformal field theories with CY target spaces at Gepner points \cite{Cordova:2023qei}. 
Our purpose is to reveal coupling selection rules of matter fields on CY compactifications with large volume regime in the context of four-dimensional low-energy effective field theory derived from heterotic string theory with standard embedding. 
As concrete examples, we consider complete intersection Calabi-Yau threefolds (CICYs) \cite{Candelas:1987kf,Candelas:1987du}. 
We classify coupling selection rules of chiral matter fields on CICYs.
Those coupling selection rules can not be understood by group theoretic rules.
We show that those coupling selection rules can be understood by combinations of essential fusion rules.

The paper is organized as follows. 
After briefly reviewing CICYs in Sec. \ref{sec:CICY}, we classify coupling selection rules of chiral matters in heterotic string theory on CICYs up to $h^{1,1}=5$ in Sec. \ref{eq:selection-rule}. 
Then, an underlying structure of coupling selection rules is discussed in Sec. \ref{sec:more}. 
Sec. \ref{sec:con} is devoted to the conclusions. In Appendix \ref{app:gauging}, we review the $\mathbb{Z}_2$ gauging of $\mathbb{Z}_M$ symmetries and extend them. 
The coupling selection rules on CICYs with $h^{1,1}=4$ and $5$ are respectively summarized in Appendices \ref{app:h11=4} and \ref{app:h11=5}.

\section{Compactifications on Complete Intersection Calabi-Yau}
\label{sec:CICY}

Here, we give a review of CY compactifications, in particular, CICYs classified in Refs.~\cite{Candelas:1987kf,Candelas:1987du,Green:1987cr,He:1990pg,Gagnon:1994ek}, 
which are defined in the ambient space such as a product of complex projective spaces: $\mathbb{P}^{n_1}\times\cdots\mathbb{P}^{n_m}$. 
The CICYs are specified by the following configuration matrix:
\begin{align}
\begin{matrix}
\mathbb{P}^{n_1}\\
\mathbb{P}^{n_2}\\
\vdots\\
\mathbb{P}^{n_m}\\
\end{matrix}
\begin{bmatrix}
q_1^1 & q_2^1 & \cdots & q_R^1\\
q_1^2 & q_2^2 & \cdots & q_R^2\\
\vdots & \vdots & \ddots & \vdots\\
q_1^m & q_2^m & \cdots & q_R^m\\
\end{bmatrix},
\label{eq:conf_matrix}
\end{align}
where $q_r^l$ ($r=1,...,R$, $l=1,...,m$) denote the positive integers, specifying a 
$R$ number of multi-degree of homogeneous polynomials on the ambient spaces $\mathbb{P}^{n_1}\times\cdots\mathbb{P}^{n_m}$. 
Then, the CICY is realized at a common zero locus of $R$ number of polynomials under 
the following conditions:
\begin{align}
    \sum_{r=1}^R q_r^l &= n_l +1\quad (\forall l),
    \nonumber\\
    \sum_{l=1}^m n_l &= R + 3,
\end{align}
where the first and second conditions respectively correspond to a vanishing first Chern class of the tangent bundle 
and a complex three-dimensional manifold. 
It was known that there exist a total of 7890 CICYs, but some of which enjoy the same topological properties \cite{Anderson:2008uw}. 
In this paper, we analyze the so-called favorable CICYs where the second cohomology of CICY descends from that of the ambient space.

When we specify the configuration matrix \eqref{eq:conf_matrix}, one can determine the topological data of 
CICYs, i.e., the second Chern number, the Euler number, the intersection number $\kappa_{abc}$ ($a,b,c=1,...,h^{1,1}$), and the number of K\"ahler moduli $h^{1,1}$ and complex structure moduli $h^{2,1}$ appearing in the four-dimensional effective action. 
The interaction of moduli fields is described by the prepotential:
\begin{align}
\label{eq:prepotential}
    {\cal F}=\frac{1}{6}\kappa_{abc}t_a t_b t_c,
\end{align}
where we focus on the large volume regime of K\"ahler moduli $t_a$.\footnote{One can obtain the same structure in the complex structure moduli sector as well.} 
Note that the intersection numbers $\kappa_{abc}$ are symmetric under permutations of indices, $a,b,c$. 
Hence, a selection rule of intersection numbers determines the interaction of moduli fields as well as matter fields in the four-dimensional effective action. 
As a concrete example, let us consider heterotic string theory on Calabi-Yau threefolds with standard embedding, 
where the gauge group is broken down to $E_8\rightarrow E_6 \times SU(3)$ or $SO(32) \rightarrow SO(28) \times SU(3)$. 
Since the $SU(3)$ gauge bundle is identified with the holomorphic tangent bundle of CY, the K\"ahler moduli and the complex structure moduli are respectively identified with the fundamental and anti-fundamental representations of, e.g., $E_6$. Hence, the selection rules of moduli fields are also identified with these of matter fields. 
In the following section, we analyze the coupling selection rule of matter fields on each CICY 
where we refer to its number in order to identify CICYs, e.g., CICY-1, following Ref. \cite{CICY}.

\section{Classification of coupling selection rules}
\label{eq:selection-rule}

As commented in Sec. \ref{sec:CICY}, there is a one-to-one correspondence between the moduli $t_a$ and chiral matter fields $A_a$ in heterotic string theory on CY compactifications.
For example, in the standard embedding of $E_6$ theory, the $h^{1,1}$ moduli $t_a$  correspond to 
the ${\bf 27}$ chiral matter fields $A_a$, while the $h^{2,1}$ moduli correspond to the $\overline {\bf  27}$ chiral matter fields.
Here, we focus on chiral matter fields $A_a$ corresponding to the 
$h^{1,1}$ moduli.

Given the prepotential of the moduli as Eq.(\ref{eq:prepotential}), 
the superpotential of matter fields $A_a$ can be written by 
\begin{align}
    W=\kappa_{abc}A_aA_bA_c.
\end{align}
Thus, intersection numbers $\kappa_{abc}$ provide us with coupling selection rules.\footnote{Traditional flavor symmetries were discussed in Refs.~\cite{Ishiguro:2021ccl,Ishiguro:2021drk} in the context of the standard embedding.} 
Moreover, chiral matter fields $A_a$ correspond to vertex operators $V_a$ in world-sheet conformal field theory.
(See for vertex operators of CY compactifications Ref.~\cite{Dixon:1989fj}.)
The coupling selection rules of chiral matter fields $A_a$ are 
written as the following fusion rule:
\begin{align}
    (V_a V_b)=\kappa_{abc}V_c,
\end{align}
by the vertex operators in the world-sheet conformal field theory.
That is a chiral ring structure.

CICYs with $h^{1,1}=1$ lead to the simple fusion rule 
as $(V_1 V_1)=V_1$ up to $\kappa_{111} \neq 0$.
We start with CICYs with $h^{1,1}=2$.

\subsection{CICYs with $h^{1,1}=2$}

Here, we study the fusion rules of the CICYs with $h^{1,1}=2$ whose number is 36. Here, we concentrate on coupling selection rules of allowing couplings $\kappa_{abc}\neq 0$ and forbidden couplings $\kappa_{abc}=0$, without paying attention to values of allowed couplings $\kappa_{abc} \neq 0$ at this stage. 
Judging from vanishing $\kappa_{abc}$, we can classify 
all the CICYs with $h^{1,1}=2$ to three types as shown in Table \ref{tab:h11=2}.
\begin{table}[H]
    \centering
       \caption{Types of prepotential for CICYs with $h^{1,1}=2$. Vanishing $\kappa_{abc}=0$ are shown.}
       \label{tab:h11=2}
    \begin{tabular}{|c||c|}
    \hline
    Type & $\kappa_{abc}=0$ \\ \hline \hline 
    1 & none \\
      \hline
      2 & $\kappa_{111}=0$ \\
      \hline
      3 & $\kappa_{111}=\kappa_{112}=0$ \\
      \hline
    \end{tabular}
  \end{table}
Concretely, the CICYs with $h^{1,1}=2$ are classified as 
\begin{align}
    &\mathrm{Type}\,1:\{ {\rm CICY-}7644, 7726, 7759, 7761, 7799, 7809, 7863  \},\notag\\
        &\mathrm{Type}\,2:\{ {\rm CICY-}7643, 7668, 7725, 7758, 7807, 7808, 7821, 7833, 7844, 7853, 7868, 7883, 7884  \},  \notag \\
            &\mathrm{Type}\,3:\{ {\rm CICY-} 7806, 7816, 7817, 7819, 7822, 7823, 7840, 7858, 7867, 7869, 7873, 7882, 7885, \notag 
            \\ &~~~~~~~~~~~~~~~~~~~~~~~7886, 7887, 7888  \}  \notag,
\end{align}
depending on the vanishing entries of the intersection numbers, as shown in Table \ref{tab:h11=2}.
Here, we call them CICY-\# following Ref.~\cite{CICY}.

For Type 1, all couplings are allowed.
Coupling selection rules are trivial.
Next, we study Type 3.
Its prepotential is written by 
\begin{align}
    {\cal F}=2\kappa_{122}t_1t_2^2+\kappa_{222}t_2^3,
\end{align}
where $\kappa_{122},\kappa_{222}\neq 0$.
That leads to the following fusion rules of $V_a$:
\begin{align}
\label{eq:FR-h11=2-2}
    (V_1V_1)=0, \qquad (V_1V_2)=V_2, \qquad (V_2V_2)=V_1+V_2,
\end{align}
up to coefficients.

The above coupling selection rules lead to a definite Yukawa texture of chiral matter fields $A_a$.
In this model, there are two generations of matter fields $A_1$ and $A_2$.
The Higgs field can correspond to either $A_1$ or $A_2$.
When the Higgs field corresponds to $A_1$, we obtain the following Yukawa texture:
\begin{align}
    Y_{ab}=
    \begin{pmatrix}
        0 & 0 \\ 0 & *
    \end{pmatrix}.
\end{align}
Here and in what follows, the asterisk symbol $*$ denotes non-vanishing entries.
This matrix has rank 1.
On the other hand, when the Higgs field corresponds to $A_2$, we obtain the following Yukawa texture:
\begin{align}
    Y_{ab}=
    \begin{pmatrix}
        0 & * \\ * & *
    \end{pmatrix}.
\end{align}
This texture can also be derived by a gauging of the $\mathbb Z_2$ outer automorphism of $\mathbb Z_3$ symmetry \cite{Kobayashi:2024cvp,Kobayashi:2025znw}, namely the Fibonacci fusion rules.

Type 2 of CICYs with $h^{1,1}=2$ leads to the following prepotential:
\begin{align}
    {\cal F}=2\kappa_{112}t_1^2t_2+2\kappa_{122}t_1t_2^2+\kappa_{222}t_2^3,
\end{align}
where $\kappa_{112},\kappa_{122}, \kappa_{222},\neq 0$.
That leads to the following fusion rules of $V_a$:
\begin{align}
\label{eq:FR-h11=2-3}
    (V_1V_1)=V_2, \qquad (V_1V_2)=V_1+V_2, \qquad (V_2V_2)=V_1+V_2,
\end{align}
up to coefficients.
The fusion rule (\ref{eq:FR-h11=2-3}) is understood by 
$\mathbb Z^{(1)}_2$ gauging of $\mathbb Z_3$ symmetry as shown in Appendix \ref{app:gauging}.

The above coupling selection rules lead to a definite Yukawa texture of chiral matter fields $A_a$.
When the Higgs field corresponds to $A_2$, we obtain the following Yukawa texture:
\begin{align}
    Y_{ab}=
    \begin{pmatrix}
        0 & * \\ * & *
    \end{pmatrix}.
\end{align}
This is the same as the matrix, which is obtained in Type 3, when the Higgs field corresponds to $A_2$.
On the other hand, when the Higgs field corresponds to $A_1$, 
the Yukawa texture is trivial, and all entries are allowed by 
these selection rules.

We have obtained three types of fusion rules leading to the selection rules of allowed couplings among chiral matter fields.
Its topological difference is clear.
We denote a real basis of harmonic (1,1)-forms by $e_a$ $(a=1,2)$ corresponding to the moduli $t_a$ and chiral matter fields $A_a$.
For Type 1, $e_1$ by itself can make six dimensional form 
\begin{align}
    \int e_1 \wedge e_1 \wedge e_1 \neq 0,
\end{align}
on the CY space.
For Type 2, $e_1$ by itself can not make six dimensional form
\begin{align}
    \int e_1 \wedge e_1 \wedge e_1 = 0.
\end{align}
Then, the self coupling $A_1^3$ vanishes on such CY space.
In addition, for Type 3, we have 
\begin{align}
    \int e_1 \wedge e_1 \wedge e_a = 0,
\end{align}
where $a=1,2$.
The CICYs with $h^{1,1}=2$ can have only three possible topological structures.

\subsection{CICYs with $h^{1,1}=3$ and $4$}

Here, we study the coupling selection rules of the CICYs with $h^{1,1}=3$ whose number is 155. 
Similarly, judging from vanishing $\kappa_{abc}$, we can classify all the CICYs with $h^{1,1}=3$ to eleven types as shown in Table \ref{tab:h11=3}, where vanishing $\kappa_{abc}$ are shown.
Concrete CICYs with $h^{1,1}=3$ are shown in Table \ref{tab:type-h11=3} as CICY-\# following \cite{CICY}.

\begin{table}[H]
    \centering
    \caption{Types of prepotential for CICYs with $h^{1,1}=3$. Vanishing $\kappa_{abc}$ are shown.}
    \label{tab:h11=3}
    \begin{tabular}{|c||c|}
    \hline
    Type & $\kappa_{abc}=0$ \\ \hline \hline
      1 & $\kappa_{111}, \kappa_{222}, \kappa_{333}$ \\
      \hline
       2 & $\kappa_{111}, \kappa_{222}$ \\
      \hline
      3 & $\kappa_{111}, \kappa_{112}, \kappa_{113}$ \\
      \hline
       4 & $\kappa_{111}, \kappa_{112}, \kappa_{113}, \kappa_{222}$ \\
      \hline
       5 & $\kappa_{111}$ \\
      \hline
       6 & none \\
      \hline
       7 & $\kappa_{111}, \kappa_{112}, \kappa_{113} ,\kappa_{122} ,\kappa_{222} ,\kappa_{223}$ \\
      \hline
       8 & $\kappa_{111}, \kappa_{112}, \kappa_{113} ,\kappa_{122} ,\kappa_{222}$ \\
      \hline
       9 & $\kappa_{111}, \kappa_{112}, \kappa_{113}, \kappa_{222}, \kappa_{333}$ \\
      \hline
       10 & $\kappa_{111}, \kappa_{112}, \kappa_{113}, \kappa_{122} ,\kappa_{222}, \kappa_{333}$ \\
      \hline
       11 & $\kappa_{111}, \kappa_{112}, \kappa_{113}, \kappa_{122}, \kappa_{222}, \kappa_{223}, \kappa_{333}$ \\
      \hline
    \end{tabular}
  \end{table}

We can write the fusion rules.
For example, in Type 1, we have the following fusion rules:
\begin{align}
    (V_aV_a)=\sum_{b\neq a}V_b, \qquad (V_aV_{b(\neq a)})=\sum_c V_c.
\end{align}
Similarly, we can write down the fusion rules for other types.

\newpage

    \begin{longtable}[h]{|c||p{12cm}|}
     \caption{Types of concrete CICYs with $h^{1,1}=3$.}  \label{tab:type-h11=3}\\
        \hline
        Type & \multicolumn{1}{c|}{Number of CICY}\\
        \hline
        \addtocounter{c2}{1}
        \arabic{c2} &
        $ 5299 ,$ 
        $ 6971 ,$ 
        $ 7580 ,$ 
        $ 7581 ,$ 
        $ 7669 ,$ 
        $ 7729 ,$ 
        $ 7846 $ 
        \\
        \hline
        \addtocounter{c2}{1}
        \arabic{c2} &
        $ 6220 ,$ 
        $ 6555 ,$ 
        $ 6827 ,$ 
        $ 6972 ,$ 
        $ 7143 ,$ 
        $ 7235 ,$ 
        $ 7240 ,$ 
        $ 7365 ,$ 
        $ 7366 ,$ 
        $ 7369 ,$ 
        $ 7486 ,$ 
        $ 7534 ,$ 
        $ 7583 ,$ 
        $ 7612 ,$ 
        $ 7646 ,$ 
        $ 7698 ,$ 
        $ 7762 ,$ 
        $ 7791 $ 
        \\
        \hline
        \addtocounter{c2}{1}
        \arabic{c2} &
        $ 6771 ,$ 
        $ 7036 ,$ 
        $ 7208 ,$ 
        $ 7530 ,$ 
        $ 7563 ,$ 
        $ 7566 ,$ 
        $ 7571 ,$ 
        $ 7578 ,$ 
        $ 7588 ,$ 
        $ 7626 ,$ 
        $ 7631 ,$ 
        $ 7635 ,$ 
        $ 7636 ,$ 
        $ 7638 ,$ 
        $ 7647 ,$ 
        $ 7648 ,$ 
        $ 7679 ,$ 
        $ 7717 ,$ 
        $ 7721 ,$ 
        $ 7734 ,$ 
        $ 7747 ,$ 
        $ 7781 ,$ 
        $ 7842 $ 
        \\
        \hline
        \addtocounter{c2}{1}
        \arabic{c2} &
        $ 7069 ,$ 
        $ 7316 ,$ 
        $ 7317 ,$ 
        $ 7452 ,$ 
        $ 7464 ,$ 
        $ 7485 ,$ 
        $ 7556 ,$ 
        $ 7558 ,$ 
        $ 7561 ,$ 
        $ 7562 ,$ 
        $ 7570 ,$ 
        $ 7584 ,$ 
        $ 7585 ,$ 
        $ 7587 ,$ 
        $ 7610 ,$ 
        $ 7627 ,$ 
        $ 7645 ,$ 
        $ 7676 ,$ 
        $ 7697 ,$ 
        $ 7710 ,$ 
        $ 7711 ,$ 
        $ 7720 ,$ 
        $ 7730 ,$ 
        $ 7752 ,$ 
        $ 7755 ,$ 
        $ 7763 ,$ 
        $ 7798 ,$ 
        $ 7802 ,$ 
        $ 7824 ,$ 
        $ 7843 ,$ 
        $ 7847 ,$ 
        $ 7854 $ 
        \\
        \hline
        \addtocounter{c2}{1}
        \arabic{c2} &
        $ 7071 ,$ 
        $ 7144 ,$ 
        $ 7237 ,$ 
        $ 7370 ,$ 
        $ 7488 ,$ 
        $ 7586 $ 
        \\
        \hline
        \addtocounter{c2}{1}
        \arabic{c2} &
        $ 7242 $ 
        \\
        \hline
        \addtocounter{c2}{1}
        \arabic{c2} &
        $ 7450 ,$ 
        $ 7481 ,$ 
        $ 7484 ,$ 
        $ 7555 ,$ 
        $ 7560 ,$ 
        $ 7579 ,$ 
        $ 7661 ,$ 
        $ 7662 ,$ 
        $ 7677 ,$ 
        $ 7694 ,$ 
        $ 7707 ,$ 
        $ 7714 ,$ 
        $ 7735 ,$ 
        $ 7745 ,$ 
        $ 7746 ,$ 
        $ 7753 ,$ 
        $ 7760 ,$ 
        $ 7769 ,$ 
        $ 7776 ,$ 
        $ 7779 ,$ 
        $ 7780 ,$ 
        $ 7788 ,$ 
        $ 7789 ,$ 
        $ 7792 ,$ 
        $ 7795 ,$ 
        $ 7797 ,$ 
        $ 7812 ,$ 
        $ 7834 ,$ 
        $ 7836 ,$ 
        $ 7841 ,$ 
        $ 7845 ,$ 
        $ 7848 ,$ 
        $ 7851 ,$ 
        $ 7865 ,$ 
        $ 7871 ,$ 
        $ 7872 ,$ 
        $ 7874 ,$ 
        $ 7877 ,$ 
        $ 7881 $ 
        \\
        \hline
        \addtocounter{c2}{1}
        \arabic{c2} &
        $ 7465 ,$ 
        $ 7466 ,$ 
        $ 7565 ,$ 
        $ 7576 ,$ 
        $ 7577 ,$ 
        $ 7637 ,$ 
        $ 7678 ,$ 
        $ 7680 ,$ 
        $ 7712 ,$ 
        $ 7713 ,$ 
        $ 7756 ,$ 
        $ 7774 ,$ 
        $ 7782 ,$ 
        $ 7783 ,$ 
        $ 7787 ,$ 
        $ 7801 ,$ 
        $ 7804 ,$ 
        $ 7832 ,$ 
        $ 7838 ,$ 
        $ 7855 ,$ 
        $ 7866 ,$ 
        $ 7876 $ 
        \\
        \hline
        \addtocounter{c2}{1}
        \arabic{c2} &
        $ 7708 ,$ 
        $ 7727 ,$ 
        $ 7728 ,$ 
        $ 7831 ,$ 
        $ 7870 $ 
        \\
        \hline
        \addtocounter{c2}{1}
        \arabic{c2} &
        $ 7875 $ 
        \\
        \hline
        \addtocounter{c2}{1}
        \arabic{c2} &
        $ 7880 $ 
        \\
        \hline
    \end{longtable}

Also, we can obtain allowed Yukawa couplings of chiral matter fields $A_a$, leading to the possible texture of three generations $(A_1,A_2,A_3)$ when we assume that the Higgs field $H$ is originated from either $A_1$, $A_2$, or $A_3$.
Results are shown in Table \ref{tab:h11=3-texture}.
Some matrices have vanishing determinants.

    \begin{table}[H]
        \centering
        \caption{Yukawa textures from CICYs with $h^{1,1}=3$.}
        \label{tab:h11=3-texture}
          \begin{tabular}{|c|c|c|c|}
            \hline
            Type \textbackslash $H$ & $A_1$ & $A_2$ & $A_3$\\
            \hline \hline
            \addtocounter{c}{1}
            Type \arabic{c} 
            & $\pmqty{
                 0  & 
                 *  & 
                 * \\ 
                 *  & 
                 *  & 
                 * \\ 
                 *  & 
                 *  & 
                 * \\ 
            }$
            & $\pmqty{
                 *  & 
                 *  & 
                 * \\ 
                 *  & 
                 0  & 
                 * \\ 
                 *  & 
                 *  & 
                 * \\ 
            }$
            & $\pmqty{
                 *  & 
                 *  & 
                 * \\ 
                 *  & 
                 *  & 
                 * \\ 
                 *  & 
                 *  & 
                 0 \\ 
            }$
            \\
            \hline
            \addtocounter{c}{1}
            Type \arabic{c} 
            & $\pmqty{
                 0  & 
                 *  & 
                 * \\ 
                 *  & 
                 *  & 
                 * \\ 
                 *  & 
                 *  & 
                 * \\ 
            }$
            & $\pmqty{
                 *  & 
                 *  & 
                 * \\ 
                 *  & 
                 0  & 
                 * \\ 
                 *  & 
                 *  & 
                 * \\ 
            }$
            & $\pmqty{
                 *  & 
                 *  & 
                 * \\ 
                 *  & 
                 *  & 
                 * \\ 
                 *  & 
                 *  & 
                 * \\ 
            }$
            \\
            \hline
            \addtocounter{c}{1}
            Type \arabic{c} 
            & $\pmqty{
                 0  & 
                 0  & 
                 0 \\ 
                 0  & 
                 *  & 
                 * \\ 
                 0  & 
                 *  & 
                 * \\ 
            }$
            & $\pmqty{
                 0  & 
                 *  & 
                 * \\ 
                 *  & 
                 *  & 
                 * \\ 
                 *  & 
                 *  & 
                 * \\ 
            }$
            & $\pmqty{
                 0  & 
                 *  & 
                 * \\ 
                 *  & 
                 *  & 
                 * \\ 
                 *  & 
                 *  & 
                 * \\ 
            }$
            \\
            \hline
            \addtocounter{c}{1}
            Type \arabic{c} 
            & $\pmqty{
                 0  & 
                 0  & 
                 0 \\ 
                 0  & 
                 *  & 
                 * \\ 
                 0  & 
                 *  & 
                 * \\ 
            }$
            & $\pmqty{
                 0  & 
                 *  & 
                 * \\ 
                 *  & 
                 0  & 
                 * \\ 
                 *  & 
                 *  & 
                 * \\ 
            }$
            & $\pmqty{
                 0  & 
                 *  & 
                 * \\ 
                 *  & 
                 *  & 
                 * \\ 
                 *  & 
                 *  & 
                 * \\ 
            }$
            \\
            \hline
            \addtocounter{c}{1}
            Type \arabic{c} 
            & $\pmqty{
                 0  & 
                 *  & 
                 * \\ 
                 *  & 
                 *  & 
                 * \\ 
                 *  & 
                 *  & 
                 * \\ 
            }$
            & $\pmqty{
                 *  & 
                 *  & 
                 * \\ 
                 *  & 
                 *  & 
                 * \\ 
                 *  & 
                 *  & 
                 * \\ 
            }$
            & $\pmqty{
                 *  & 
                 *  & 
                 * \\ 
                 *  & 
                 *  & 
                 * \\ 
                 *  & 
                 *  & 
                 * \\ 
            }$
            \\
            \hline
            \addtocounter{c}{1}
            Type \arabic{c} 
            & $\pmqty{
                 *  & 
                 *  & 
                 * \\ 
                 *  & 
                 *  & 
                 * \\ 
                 *  & 
                 *  & 
                 * \\ 
            }$
            & $\pmqty{
                 *  & 
                 *  & 
                 * \\ 
                 *  & 
                 *  & 
                 * \\ 
                 *  & 
                 *  & 
                 * \\ 
            }$
            & $\pmqty{
                 *  & 
                 *  & 
                 * \\ 
                 *  & 
                 *  & 
                 * \\ 
                 *  & 
                 *  & 
                 * \\ 
            }$
            \\
            \hline
            \addtocounter{c}{1}
            Type \arabic{c} 
            & $\pmqty{
                 0  & 
                 0  & 
                 0 \\ 
                 0  & 
                 0  & 
                 * \\ 
                 0  & 
                 *  & 
                 * \\ 
            }$
            & $\pmqty{
                 0  & 
                 0  & 
                 * \\ 
                 0  & 
                 0  & 
                 0 \\ 
                 *  & 
                 0  & 
                 * \\ 
            }$
            & $\pmqty{
                 0  & 
                 *  & 
                 * \\ 
                 *  & 
                 0  & 
                 * \\ 
                 *  & 
                 *  & 
                 * \\ 
            }$
            \\
            \hline
            \addtocounter{c}{1}
            Type \arabic{c} 
            & $\pmqty{
                 0  & 
                 0  & 
                 0 \\ 
                 0  & 
                 0  & 
                 * \\ 
                 0  & 
                 *  & 
                 * \\ 
            }$
            & $\pmqty{
                 0  & 
                 0  & 
                 * \\ 
                 0  & 
                 0  & 
                 * \\ 
                 *  & 
                 *  & 
                 * \\ 
            }$
            & $\pmqty{
                 0  & 
                 *  & 
                 * \\ 
                 *  & 
                 *  & 
                 * \\ 
                 *  & 
                 *  & 
                 * \\ 
            }$
            \\
            \hline
            \addtocounter{c}{1}
            Type \arabic{c} 
            & $\pmqty{
                 0  & 
                 0  & 
                 0 \\ 
                 0  & 
                 *  & 
                 * \\ 
                 0  & 
                 *  & 
                 * \\ 
            }$
            & $\pmqty{
                 0  & 
                 *  & 
                 * \\ 
                 *  & 
                 0  & 
                 * \\ 
                 *  & 
                 *  & 
                 * \\ 
            }$
            & $\pmqty{
                 0  & 
                 *  & 
                 * \\ 
                 *  & 
                 *  & 
                 * \\ 
                 *  & 
                 *  & 
                 0 \\ 
            }$
            \\
            \hline
            \addtocounter{c}{1}
            Type \arabic{c} 
            & $\pmqty{
                 0  & 
                 0  & 
                 0 \\ 
                 0  & 
                 0  & 
                 * \\ 
                 0  & 
                 *  & 
                 * \\ 
            }$
            & $\pmqty{
                 0  & 
                 0  & 
                 * \\ 
                 0  & 
                 0  & 
                 * \\ 
                 *  & 
                 *  & 
                 * \\ 
            }$
            & $\pmqty{
                 0  & 
                 *  & 
                 * \\ 
                 *  & 
                 *  & 
                 * \\ 
                 *  & 
                 *  & 
                 0 \\ 
            }$
            \\
            \hline
            \addtocounter{c}{1}
            Type \arabic{c} 
            & $\pmqty{
                 0  & 
                 0  & 
                 0 \\ 
                 0  & 
                 0  & 
                 * \\ 
                 0  & 
                 *  & 
                 * \\ 
            }$
            & $\pmqty{
                 0  & 
                 0  & 
                 * \\ 
                 0  & 
                 0  & 
                 0 \\ 
                 *  & 
                 0  & 
                 * \\ 
            }$
            & $\pmqty{
                 0  & 
                 *  & 
                 * \\ 
                 *  & 
                 0  & 
                 * \\ 
                 *  & 
                 *  & 
                 0 \\ 
            }$
            \\
            \hline
        \end{tabular}
     \end{table}

Similarly, we can classify the prepotential of CICYs with $h^{1,1} =4$ whose number is 425. 
Results are shown in Table \ref{tab:h11=4-k}.
Also, we can write possible texture of four generations $(A_1,A_2,A_3,A_4)$ when we assume that the Higgs field is originated from either $A_1$, $A_2$, $A_3$, or $A_4$.
Results are shown in Appendix \ref{app:h11=4}.

\newpage

 \begin{longtable}[H]{|c||p{12cm}|}
 \caption{Types of prepotential for CICYs with $h^{1,1}=4$. Vanishing $\kappa_{abc}$ are shown.}
\label{tab:h11=4-k}\\
\hline
Type & \multicolumn{1}{c|}{$\kappa_{abc}=0$} \\
\hline
1 & 
$\kappa_{ 1 1 1 }$,
$\kappa_{ 1 1 2 }$,
$\kappa_{ 1 1 3 }$,
$\kappa_{ 1 1 4 }$,
$\kappa_{ 1 2 2 }$,
$\kappa_{ 2 2 2 }$,
$\kappa_{ 3 3 3 }$,
$\kappa_{ 4 4 4 }$
\\ \hline
2 & 
$\kappa_{ 1 1 1 }$,
$\kappa_{ 2 2 2 }$,
$\kappa_{ 3 3 3 }$
\\ \hline
3 & 
$\kappa_{ 1 1 1 }$,
$\kappa_{ 1 1 2 }$,
$\kappa_{ 1 1 3 }$,
$\kappa_{ 1 1 4 }$,
$\kappa_{ 1 2 2 }$,
$\kappa_{ 2 2 2 }$,
$\kappa_{ 2 2 3 }$,
$\kappa_{ 2 2 4 }$,
$\kappa_{ 3 3 3 }$
\\ \hline
4 & 
$\kappa_{ 1 1 1 }$,
$\kappa_{ 1 1 2 }$,
$\kappa_{ 1 1 3 }$,
$\kappa_{ 1 1 4 }$,
$\kappa_{ 1 2 2 }$,
$\kappa_{ 2 2 2 }$,
$\kappa_{ 3 3 3 }$
\\ \hline
5 & 
$\kappa_{ 1 1 1 }$,
$\kappa_{ 2 2 2 }$,
$\kappa_{ 3 3 3 }$,
$\kappa_{ 4 4 4 }$
\\ \hline
6 & 
$\kappa_{ 1 1 1 }$,
$\kappa_{ 1 1 2 }$,
$\kappa_{ 1 1 3 }$,
$\kappa_{ 1 1 4 }$,
$\kappa_{ 1 3 3 }$,
$\kappa_{ 2 2 2 }$,
$\kappa_{ 3 3 3 }$,
$\kappa_{ 4 4 4 }$
\\ \hline
7 & 
$\kappa_{ 1 1 1 }$,
$\kappa_{ 1 1 2 }$,
$\kappa_{ 1 1 3 }$,
$\kappa_{ 1 1 4 }$,
$\kappa_{ 2 2 2 }$,
$\kappa_{ 3 3 3 }$
\\ \hline
8 & 
$\kappa_{ 1 1 1 }$,
$\kappa_{ 1 1 2 }$,
$\kappa_{ 1 1 3 }$,
$\kappa_{ 1 1 4 }$,
$\kappa_{ 1 3 3 }$,
$\kappa_{ 2 2 2 }$,
$\kappa_{ 3 3 3 }$
\\ \hline
9 & 
$\kappa_{ 1 1 1 }$,
$\kappa_{ 2 2 2 }$
\\ \hline
10 & 
$\kappa_{ 1 1 1 }$,
$\kappa_{ 1 1 2 }$,
$\kappa_{ 1 1 3 }$,
$\kappa_{ 1 1 4 }$,
$\kappa_{ 2 2 2 }$,
$\kappa_{ 3 3 3 }$,
$\kappa_{ 4 4 4 }$
\\ \hline
11 & 
$\kappa_{ 1 1 1 }$,
$\kappa_{ 1 1 2 }$,
$\kappa_{ 1 1 3 }$,
$\kappa_{ 1 1 4 }$,
$\kappa_{ 1 2 2 }$,
$\kappa_{ 2 2 2 }$,
$\kappa_{ 2 2 3 }$,
$\kappa_{ 2 2 4 }$,
$\kappa_{ 3 3 3 }$,
$\kappa_{ 4 4 4 }$
\\ \hline
12 & 
$\kappa_{ 1 1 1 }$,
$\kappa_{ 1 1 2 }$,
$\kappa_{ 1 1 3 }$,
$\kappa_{ 1 1 4 }$,
$\kappa_{ 2 2 2 }$
\\ \hline
13 & 
$\kappa_{ 1 1 1 }$,
$\kappa_{ 1 1 2 }$,
$\kappa_{ 1 1 3 }$,
$\kappa_{ 1 1 4 }$,
$\kappa_{ 1 2 2 }$,
$\kappa_{ 2 2 2 }$
\\ \hline
14 & 
$\kappa_{ 1 1 1 }$,
$\kappa_{ 1 1 2 }$,
$\kappa_{ 1 1 3 }$,
$\kappa_{ 1 1 4 }$,
$\kappa_{ 1 2 2 }$,
$\kappa_{ 2 2 2 }$,
$\kappa_{ 2 2 3 }$,
$\kappa_{ 2 2 4 }$
\\ \hline
15 & 
$\kappa_{ 1 1 1 }$,
$\kappa_{ 1 1 2 }$,
$\kappa_{ 1 1 3 }$,
$\kappa_{ 1 1 4 }$,
$\kappa_{ 1 2 2 }$,
$\kappa_{ 1 3 3 }$,
$\kappa_{ 2 2 2 }$,
$\kappa_{ 2 2 3 }$,
$\kappa_{ 2 2 4 }$,
$\kappa_{ 2 3 3 }$,
$\kappa_{ 3 3 3 }$,
$\kappa_{ 3 3 4 }$
\\ \hline
16 & 
$\kappa_{ 1 1 1 }$,
$\kappa_{ 1 1 2 }$,
$\kappa_{ 1 1 3 }$,
$\kappa_{ 1 1 4 }$
\\ \hline
17 & 
$\kappa_{ 1 1 1 }$,
$\kappa_{ 1 1 2 }$,
$\kappa_{ 1 1 3 }$,
$\kappa_{ 1 1 4 }$,
$\kappa_{ 1 2 2 }$,
$\kappa_{ 1 3 3 }$,
$\kappa_{ 2 2 2 }$,
$\kappa_{ 2 2 3 }$,
$\kappa_{ 2 2 4 }$,
$\kappa_{ 3 3 3 }$
\\ \hline
18 & 
$\kappa_{ 1 1 1 }$,
$\kappa_{ 1 1 2 }$,
$\kappa_{ 1 1 3 }$,
$\kappa_{ 1 1 4 }$,
$\kappa_{ 1 4 4 }$,
$\kappa_{ 2 2 2 }$,
$\kappa_{ 3 3 3 }$,
$\kappa_{ 4 4 4 }$
\\ \hline
19 & 
$\kappa_{ 1 1 1 }$,
$\kappa_{ 1 1 2 }$,
$\kappa_{ 1 1 3 }$,
$\kappa_{ 1 1 4 }$,
$\kappa_{ 1 2 2 }$,
$\kappa_{ 1 3 3 }$,
$\kappa_{ 2 2 2 }$,
$\kappa_{ 3 3 3 }$
\\ \hline
20 & 
$\kappa_{ 1 1 1 }$,
$\kappa_{ 1 1 2 }$,
$\kappa_{ 1 1 3 }$,
$\kappa_{ 1 1 4 }$,
$\kappa_{ 1 2 2 }$,
$\kappa_{ 1 2 3 }$,
$\kappa_{ 1 3 3 }$,
$\kappa_{ 2 2 2 }$,
$\kappa_{ 2 2 3 }$,
$\kappa_{ 2 2 4 }$,
$\kappa_{ 2 3 3 }$,
$\kappa_{ 3 3 3 }$
\\ \hline
21 & 
$\kappa_{ 1 1 1 }$,
$\kappa_{ 1 1 2 }$,
$\kappa_{ 1 1 3 }$,
$\kappa_{ 1 1 4 }$,
$\kappa_{ 1 2 2 }$,
$\kappa_{ 1 3 3 }$,
$\kappa_{ 2 2 2 }$,
$\kappa_{ 2 2 3 }$,
$\kappa_{ 2 2 4 }$,
$\kappa_{ 2 3 3 }$,
$\kappa_{ 3 3 3 }$,
$\kappa_{ 3 3 4 }$,
$\kappa_{ 4 4 4 }$
\\ \hline
22 & 
$\kappa_{ 1 1 1 }$,
$\kappa_{ 1 1 2 }$,
$\kappa_{ 1 1 3 }$,
$\kappa_{ 1 1 4 }$,
$\kappa_{ 1 2 2 }$,
$\kappa_{ 1 3 3 }$,
$\kappa_{ 2 2 2 }$,
$\kappa_{ 2 2 3 }$,
$\kappa_{ 2 2 4 }$,
$\kappa_{ 3 3 3 }$,
$\kappa_{ 4 4 4 }$
\\ \hline
23 & 
$\kappa_{ 1 1 1 }$,
$\kappa_{ 1 1 2 }$,
$\kappa_{ 1 1 3 }$,
$\kappa_{ 1 1 4 }$,
$\kappa_{ 1 2 2 }$,
$\kappa_{ 1 4 4 }$,
$\kappa_{ 2 2 2 }$,
$\kappa_{ 2 2 3 }$,
$\kappa_{ 2 2 4 }$,
$\kappa_{ 3 3 3 }$,
$\kappa_{ 4 4 4 }$
\\ \hline
24 & 
$\kappa_{ 1 1 1 }$,
$\kappa_{ 1 1 2 }$,
$\kappa_{ 1 1 3 }$,
$\kappa_{ 1 1 4 }$,
$\kappa_{ 1 3 3 }$,
$\kappa_{ 1 4 4 }$,
$\kappa_{ 2 2 2 }$,
$\kappa_{ 3 3 3 }$,
$\kappa_{ 4 4 4 }$
\\ \hline
25 & 
$\kappa_{ 1 1 1 }$,
$\kappa_{ 1 1 2 }$,
$\kappa_{ 1 1 3 }$,
$\kappa_{ 1 1 4 }$,
$\kappa_{ 1 2 2 }$,
$\kappa_{ 1 3 3 }$,
$\kappa_{ 2 2 2 }$,
$\kappa_{ 3 3 3 }$,
$\kappa_{ 4 4 4 }$
\\ \hline
26 & 
$\kappa_{ 1 1 1 }$,
$\kappa_{ 1 1 2 }$,
$\kappa_{ 1 1 3 }$,
$\kappa_{ 1 1 4 }$,
$\kappa_{ 1 2 2 }$,
$\kappa_{ 2 2 2 }$,
$\kappa_{ 2 2 3 }$,
$\kappa_{ 2 2 4 }$,
$\kappa_{ 2 3 3 }$,
$\kappa_{ 3 3 3 }$
\\ \hline
27 & 
$\kappa_{ 1 1 1 }$,
$\kappa_{ 1 1 2 }$,
$\kappa_{ 1 1 3 }$,
$\kappa_{ 1 1 4 }$,
$\kappa_{ 1 2 2 }$,
$\kappa_{ 1 4 4 }$,
$\kappa_{ 2 2 2 }$,
$\kappa_{ 3 3 3 }$,
$\kappa_{ 4 4 4 }$
\\ \hline
28 & 
$\kappa_{ 1 1 1 }$,
$\kappa_{ 1 1 2 }$,
$\kappa_{ 1 1 3 }$,
$\kappa_{ 1 1 4 }$,
$\kappa_{ 1 2 2 }$,
$\kappa_{ 1 2 3 }$,
$\kappa_{ 1 3 3 }$,
$\kappa_{ 2 2 2 }$,
$\kappa_{ 2 2 3 }$,
$\kappa_{ 2 2 4 }$,
$\kappa_{ 2 3 3 }$,
$\kappa_{ 3 3 3 }$,
$\kappa_{ 4 4 4 }$
\\ \hline
29 & 
$\kappa_{ 1 1 1 }$,
$\kappa_{ 1 1 2 }$,
$\kappa_{ 1 1 3 }$,
$\kappa_{ 1 1 4 }$,
$\kappa_{ 1 2 2 }$,
$\kappa_{ 1 3 3 }$,
$\kappa_{ 1 4 4 }$,
$\kappa_{ 2 2 2 }$,
$\kappa_{ 2 2 3 }$,
$\kappa_{ 2 2 4 }$,
$\kappa_{ 2 3 3 }$,
$\kappa_{ 3 3 3 }$,
$\kappa_{ 3 3 4 }$,
$\kappa_{ 4 4 4 }$
\\ \hline
30 & 
$\kappa_{ 1 1 1 }$,
$\kappa_{ 1 1 2 }$,
$\kappa_{ 1 1 3 }$,
$\kappa_{ 1 1 4 }$,
$\kappa_{ 1 2 2 }$,
$\kappa_{ 1 3 3 }$,
$\kappa_{ 1 4 4 }$,
$\kappa_{ 2 2 2 }$,
$\kappa_{ 2 2 3 }$,
$\kappa_{ 2 2 4 }$,
$\kappa_{ 3 3 3 }$,
$\kappa_{ 4 4 4 }$
\\ \hline
31 & 
$\kappa_{ 1 1 1 }$,
$\kappa_{ 1 1 2 }$,
$\kappa_{ 1 1 3 }$,
$\kappa_{ 1 1 4 }$,
$\kappa_{ 1 2 2 }$,
$\kappa_{ 1 3 3 }$,
$\kappa_{ 1 4 4 }$,
$\kappa_{ 2 2 2 }$,
$\kappa_{ 2 2 3 }$,
$\kappa_{ 2 2 4 }$,
$\kappa_{ 2 3 3 }$,
$\kappa_{ 2 4 4 }$,
$\kappa_{ 3 3 3 }$,
$\kappa_{ 3 3 4 }$,
$\kappa_{ 3 4 4 }$,
$\kappa_{ 4 4 4 }$
\\ \hline
32 & 
$\kappa_{ 1 1 1 }$,
$\kappa_{ 1 1 2 }$,
$\kappa_{ 1 1 3 }$,
$\kappa_{ 1 1 4 }$,
$\kappa_{ 1 2 2 }$,
$\kappa_{ 1 4 4 }$,
$\kappa_{ 2 2 2 }$,
$\kappa_{ 2 2 3 }$,
$\kappa_{ 2 2 4 }$,
$\kappa_{ 2 3 3 }$,
$\kappa_{ 3 3 3 }$,
$\kappa_{ 4 4 4 }$
\\ \hline
\end{longtable}

Similarly, we can classify the prepotential for CICYs with $h^{1,1} =5$ whose number is 856. 
In Appendix \ref{app:h11=5}, we show classifications for $h^{1,1}=5$.

\section{Systematic study on fusion rules}
\label{sec:more}

We have classified coupling selection rules and shown results for $h^{1,1} \leq 5$ in Section 
\ref{eq:selection-rule} and Appendices \ref{app:h11=4} and \ref{app:h11=5}.
Here, we provide a systematic understanding of their fusion rules.

First, the coupling selection rule in CICYs with $h^{1,1}=1$ is very simple.
That is, the 3-point coupling $A_1^3$ is allowed.
We call such a symmetry S1.
There is a single element $[a_1]$ in S1, and it satisfies the following fusion rule:
\begin{align}
    {\rm S1:}~~~[a_1][a_1]=[a_1].
\end{align}

We have shown that there are three types of fusion rules for $h^{1,1}=2$.
Type 1 corresponds to the above symmetry S1, where both $A_1$ and $A_2$ correspond to $[a_1]$.
In addition, there are two non-trivial selection rules including two elements $[a_{2,1}]$ and $[a_{2,2}]$.
We define two symmetries S2-a and S2-b, which correspond to Types  2 and 3, respectively.
For the S2-a symmetry, two elements  $[a_{2,1}]$ and $[a_{2,2}]$ satisfy the following selection rules:
\begin{align}
    {\rm S2-a:}~~~[a_{2,1}][a_{2,1}]=[a_{2,2}], \qquad [a_{2,1}][a_{2,2}]=[a_{2,1}]+[a_{2,2}], \qquad [a_{2,2}][a_{2,2}]=[a_{2,1}]+[a_{2,2}].
\end{align}
For the S2-b symmetry, they satisfy 
\begin{align}
    {\rm S2-b:}~~~[b_{2,1}][b_{2,1}]=0, \qquad [b_{2,1}][b_{2,2}]=[b_{2,2}], \qquad [b_{2,2}][b_{2,2}]=[b_{2,1}]+[b_{2,2}].
\end{align}
The S2-a symmetry can be realized by $\mathbb Z^{(1)}_2$ gauging of $\mathbb Z_3$, where $[a_{2,1}]$ and $[a_{2,2}]$ correspond to $[g^1]_2$ and $[g^0]_2$ as shown in Appendix \ref{app:gauging}.

Now let us examine the coupling selection rules for CICYs with $h^{1,1}=3$.
We find that all types can be understood by combinations of S1, S2-a, S2-b except Types 8 and 10.
Type 6 is simple, and it is understood by the S1 symmetry, where all chiral fields $A_{1,2,3}$ correspond to $[a_1]$.
Type 5 is also simple, and it is understood by the S2-a symmetry, where 
$A_1$ corresponds to $[a_{2,1}]$ and the other $A_2,A_3$ correspond to $[a_{2,2}]$.
Type 2 can be understood by a combination of two S2-a symmetries, (S2-a) $\times$ (S2-a).
For the first S2-a symmetry, $A_1$ corresponds to $[a_{2,1}]$ and the other $A_2,A_3$ correspond to $[a_{2,2}]$.
For the second S2-a symmetry, $A_2$ corresponds to $[a_{2,1}]$ and the other $A_1,A_3$ correspond to $[a_{2,2}]$.
Similarly, we can understand Type 1 by a combination of three S2-a symmetries, (S2-a) $\times$(S2-a) $\times$(S2-a).
From this notation, it would be convenient to write the symmetry of Types 6, 5, and 2 as (S1) $\times$(S1) $\times$(S1), (S2-a) $\times$(S1) $\times$(S1), (S2-a) $\times$(S2-a) $\times$(S1), respectively.

On the other hand, Type 3 is understood by the symmetry (S2-b) $\times$(S1) $\times$(S1), where $A_1$ corresponds to 
$[b_{2,1}][a_1][a_1]$ and  the other $A_2,A_3$ correspond to $[b_{2,2}][a_1][a_1]$.
Similarly, Type 4 is understood by the symmetry (S2-b) $\times$(S2-a) $\times$(S1), where $A_1$ corresponds to 
$[b_{2,1}][a_{2,2}][a_1]$, $A_2$ correspond to $[b_{2,2}][a_{2,1}][a_1]$, 
and $A_3$ corresponds to  $[b_{2,2}][a_{2,2}][a_1]$.
Other types except Types 8 and 10 can be understood by combinations of S1, S2-a, S2-b symmetries.
They are shown in Table \ref{tab:h11=3-2}.

Types 8 and 10 include another symmetry of three elements.
We define the S3 symmetry, where three elements $[a_{3,i}]$ with $i=1,2,3$
satisfy the following fusion rules:
\begin{align}
\mathrm{S3:}~~~ &   [a_{3,2}][a_{3,2}]=[a_{3,3}], \quad  [a_{3,1}][a_{3,2}]=[a_{3,1}]+[a_{3,3}], \\
   & [a_{3,i}][a_{3,j}]=\sum_{k=1,2,3}[a_{3,k}] {\rm ~~for ~~other ~~combinations}. \notag
\end{align}

\begin{table}[h]
    \centering
    \caption{Symmetries for $h^{1,1}=3$.}
    \label{tab:h11=3-2}
    \begin{tabular}{|c||c|c|c|c|}
    \hline
    Type & Symmetries & $A_1$ & $A_2$ & $A_3$ \\
   \hline \hline  
       1 &  (S2-a)$\times$  (S2-a)$\times$(S2-a) & $[a_{2,1}][a_{2,2}][a_{2,2}]$ & $[a_{2,2}][a_{2,1}][a_{2,2}]$ & $[a_{2,2}][a_{2,2}][a_{2,1}]$ \\
      \hline
       2 &  (S2-a)$\times$  (S2-a)$\times$(S1) & $[a_{2,1}][a_{2,2}][a_{1}]$ & $[a_{2,2}][a_{2,1}][a_{1}]$ & $[a_{2,2}][a_{2,2}][a_{1}]$\\
      \hline
       3 &  (S2-b)$\times$  (S1)$\times$(S1) & $[b_{2,1}][a_{1}][a_{1}]$ & $[b_{2,2}][a_{1}][a_{1}]$ & $[b_{2,2}][a_{1}][a_{1}]$ \\
      \hline
       4 &  (S2-b)$\times$  (S2-a)$\times$(S1) &$[b_{2,1}][a_{2,2}][a_{1}]$ & $[b_{2,2}][a_{2,1}][a_{1}]$ & $[b_{2,2}][a_{2,2}][a_{1}]$ \\
      \hline
       5 & (S2-a)$\times$  (S1)$\times$(S1) &$[a_{2,1}][a_{1}][a_{1}]$ & $[a_{2,2}][a_{1}][a_{1}]$ & $[a_{2,2}][a_{1}][a_{1}]$  \\
      \hline
       6 & (S1)$\times$  (S1)$\times$(S1)&$[a_{1}][a_{1}][a_{1}]$ & $[a_{1}][a_{1}][a_{1}]$ & $[a_{1}][a_{1}][a_{1}]$ \\
      \hline
       7 &  (S2-b)$\times$  (S2-b)$\times$(S1) &$[b_{2,1}][b_{2,2}][a_{1}]$ & $[b_{2,2}][b_{2,1}][a_{1}]$ & $[b_{2,2}][b_{2,2}][a_{1}]$  \\
      \hline
       $8^*$ & (S2-b)$\times$  (S3)$\times$(S1) & $[b_{2,1}][a_{3,1}][a_{1}]$ & $[b_{2,2}][a_{3,2}][a_{1}]$ & $[b_{2,2}][a_{3,3}][a_{1}]$\\
      \hline
       9 &   (S2-b)$\times$  (S2-a)$\times$(S2-a)  & $[b_{2,1}][a_{2,2}][a_{2,2}]$ & $[b_{2,2}][a_{2,1}][a_{2,2}]$ & $[b_{2,2}][a_{2,2}][a_{2,1}]$ \\
      \hline
       $10^*$ & (S2-b)$\times$  (S3)$\times$(S2-a) & $[b_{2,1}][a_{3,1}][a_{2,2}]$ & $[b_{2,2}][a_{3,2}][a_{2,2}]$ & $[b_{2,2}][a_{3,3}][a_{2,1}]$\\
      \hline
       11 &  (S2-b)$\times$  (S2-b)$\times$(S2-a)  & $[b_{2,1}][b_{2,2}][a_{2,2}]$ & $[b_{2,2}][b_{2,1}][a_{2,2}]$ & $[b_{2,2}][b_{2,2}][a_{2,1}]$ \\
      \hline
    \end{tabular}
  \end{table}

Similarly,  we study symmetries for $h^{1,1}=4$.
Results are shown in Appendix \ref{app:h11=4}.
Most of the selection rules can be understood by 
fusion rules of S1, S2-a, S2-b, and S3. 
However, we need a new type of fusion rules including three elements, i.e., S4. 
These fusion rules are written by 

  \begin{align}
  \label{eq:S4}
\mathrm{S4:}~~~ 
      &[a_{3,1}][a_{3,1}] = [a_{3,1}]+[a_{3,3}], \notag\\
      &[a_{3,1}][a_{3,2}] = [a_{3,2}]+[a_{3,3}],\\
      &[a_{3,i}][a_{3,j}]=\sum_{k=1,2,3}[a_{3,k}] {\rm ~~for ~~other ~~combinations}.  \notag
\end{align}

Similarly, we can study coupling selection rules for CICYs with $h^{1,1}=5$.
For CICYs with $h^{1,1}=5$, all of the selection rules can be understood by combinations of $\{$S1, S2-a, S2-b, S3, S4$\}$. 
Here, we summarize the fusion rules, which appear to explain coupling selection rules for $h^{1,1} \leq 5$ in Table \ref{tab:summary-FR}.

\begin{table}[H]
  \centering
  \caption{Fusion rules for CICYs with $h^{1,1} = 1,2,3,4,5$.}
  \label{tab:summary-FR}
  \begin{tabular}{|c|c|}
  \hline
  \multicolumn{2}{|c|}{Fusion rules}\\
  \hline
    S1 & $[a_1][a_1]=[a_1].$ \\
    \hline
     & $[a_{2,1}][a_{2,1}]=[a_{2,2}]$,\\ 
    \multirow{1}{*}{S2-a} & $[a_{2,1}][a_{2,2}]=[a_{2,1}]+[a_{2,2}]$, \\ 
     & $[a_{2,2}][a_{2,2}]=[a_{2,1}]+[a_{2,2}]$. \\
     \hline
     & $[b_{2,1}][b_{2,1}]=0$, \\
     \multirow{1}{*}{S2-b} & $[b_{2,1}][b_{2,2}]=[b_{2,2}]$,\\
     & $[b_{2,2}][b_{2,2}]=[b_{2,1}]+[b_{2,2}].$\\
     \hline
     & $[a_{3,2}][a_{3,2}]=[a_{3,3}]$, \\
     \multirow{1}{*}{S3}& $[a_{3,1}][a_{3,2}]=[a_{3,1}]+[a_{3,3}]$, \\
     & $[a_{3,i}][a_{3,j}]=\sum_{k=1,2,3}[a_{3,k}]$.\\ 
     \hline
      &$[a_{3,1}][a_{3,1}] = [a_{3,1}]+[a_{3,3}]$,\\
      \multirow{1}{*}{S4}&$[a_{3,1}][a_{3,2}] = [a_{3,2}]+[a_{3,3}]$,\\
      &$[a_{3,i}][a_{3,j}]=\sum_{k=1,2,3}[a_{3,k}]$.\\
      \hline
  \end{tabular}
\end{table}

\section{Conclusions}
\label{sec:con}

Non-trivial selection rules associated with and without group-like symmetries have been explored in field theory and string theory. 
In this work, we focused on heterotic string theory with standard embedding from the viewpoint of four-dimensional low-energy effective field theory. 
Since the selection rules of chiral matters are originated from those of moduli fields of CY threefolds, it is expected that the CY intersection numbers lead to non-trivial selection rules. 

By classifying the selection rules on CICYs with $h^{1,1}\leq 5$, we revealed that all of the selection rules of chiral matter fields are understood by combinations of only five types of fusion rules. One of them can be realized by $\mathbb{Z}_2^{(m)}$ gauging of $\mathbb{Z}_M$ symmetry where $\mathbb{Z}_2^{(m)}$ is an outer automorphism of $\mathbb{Z}_M$. It indicates that the world sheet CFT admits the non-invertible symmetry, but it is beyond our scope and we will report this interesting direction for future work. 
Since our analysis is limited on the large volume regime of CY threefolds, it is of interest to investigate instanton corrections to the selection rules, which is left for future work. 
Furthermore, it is interesting to explore the selection rules on more broad class of CY threefolds such as the Kreuzer-Skarke database \cite{Kreuzer:2000xy}, which will be addressed in future work.

In this paper, we focused on heterotic string theory with standard embedding, but it is an interesting direction to pursue the selection rules of matter fields in other corners of string compactifications, such as non-standard embedding. Indeed, four-dimensional semi-realistic models were derived from $E_8\times E_8$ and $SO(32)$ heterotic string theories on CICYs with line bundles in e.g., Refs.~\cite{Anderson:2011ns,Anderson:2012yf,Otsuka:2018oyf,Otsuka:2020nsk}. 
We leave a comprehensive study about the selection rules of chiral matters in the future.

\acknowledgments

This work was supported in part by JSPS KAKENHI Grant Numbers JP23K03375 (T.K.)
5, JP25KJ1927 (N.S.) and JP25H01539 (H.O.).

\appendix

\section{${\mathbb Z}_2$ gauging of ${\mathbb Z}_M$ symmetries}
\label{app:gauging}

First, we briefly review ${\mathbb Z}_2$ gauging of ${\mathbb Z}_M$ symmetries, which was studied in Refs.~\cite{Kobayashi:2024yqq,Kobayashi:2024cvp}.
After that, we extend it.

We start with the $\mathbb Z_M$ symmetric theory.
We denote its generator by $g=e^{2\pi i/M}$ and 
group elements are presented by $\{g^0,g^1, g^2,\cdots,g^{M-1} \}$.
Each mode $\phi^j$ in the theory transforms 
\begin{align}
    \phi^j \to g^j\phi^j,
\end{align}
under the $\mathbb Z_M$ symmetry.
Such a $\mathbb Z_M$ symmetry can be realized 
e.g. by magnetized compactifications \cite{Abe:2009vi,Berasaluce-Gonzalez:2012abm,Marchesano:2013ega}.
On top of that, we perform geometrical $\mathbb Z_2$ orbifolding.
All of the modes $\phi^j$ are not invariant under orbifolding, but 
$\mathbb Z_2$-invariant modes can be written by \cite{Abe:2008fi}
\begin{align}
    \Phi^j = \phi^j + \phi^{M-j}. 
\end{align}
In general, these modes $\Phi^j$ have no definite $\mathbb Z_M$ charges unlike $\phi^j$.
However, they satisfy certain coupling selection rules, which are originated from the $\mathbb Z_M$ symmetry.
Such selection rules can be represented as follows.

We consider the following automorphism:
\begin{align}
\label{eq:auto}
    eg^ke^{-1}=g^k,\qquad rg^{k}r^{-1}=g^{M-k}.
\end{align}
The latter is an outer automorphism $\mathbb Z_2$, while the former is trivial. 
Hence, it corresponds to $D_M\cong \mathbb{Z}_M\rtimes \mathbb{Z}_2$. 
By gauging $\mathbb{Z}_2$, one can define the class $[g^k]$ by 
\begin{align}
    [g^k]=\{ hg^kh^{-1} | h=e,r  \}.
\end{align}
We have one-to-one correspondence between the class $[g^k]$ and the $\mathbb Z_2$-invariant mode $\Phi^k$.
The fusion rule of $[g^k]$ is obtained by
\begin{align}
    [g^k][g^{k'}]=[g^{k+k'}] + [g^{-k+k'}].
\end{align}

For example, when $M=3$, we have two classes, $[g^0]$ and $[g^1]$, where the latter includes $g^1$ and $g^2$ and the former includes only $g^0$.
They satisfy the following fusion rules:
\begin{align}
    [g^0][g^0]=[g^0],\qquad [g^0][g^1]=[g^1], \qquad 
    [g^1][g^1]=[g^0]+[g^1].
\end{align}
These fusion rules are equivalent to 
the Fibonacci fusion rules.

Now, let us extend the above $\mathbb Z_2$ gauging.
Here, the extended $\mathbb Z_2$ gauging is  denoted by 
$\mathbb Z_2^{(m)}$.
We generalize the automorphism (\ref{eq:auto}) to 
\begin{align}
    eg^ke^{-1}=g^k,\qquad r_mg^{k}r^{-1}_m=g^{M+m-k}.
\end{align}
The latter is still the $\mathbb{Z}_2$ automorphism, but it includes 
a "shift" of $\mathbb{Z}_M$ charge.
Indeed, that corresponds to a geometrical shift of $\mathbb Z_2$ orbifolding.
The orbifolding for $m=0$ corresponds to the $\mathbb Z_2$ twist around the origin, but one for $m \neq 0$ is the orbifold twist around a different point.
At any rate, we define the class $[g^k]_m$ by 
\begin{align}
    [g^k]_m=\{ hg^kh^{-1} | h=e,r_m  \},
\end{align}
for a fixed value of $m$.
Then, we arrive at a new fusion rule of $[g^k]_m$ as 
\begin{align}
    [g^k]_m[g^{k'}]_m=[g^{k+k'}]_m + [g^{m-k+k'}]_m.
\end{align}

Let us show an example with $M=3$ and $m=2$.
In this case, we have two classes: $[g^0]_2$ and $[g^1]_2$.
The class $[g^0]_2$ includes $g^0$ and $g^2$, while $[g^1]_2$ includes only $g^1$.
Their fusion rules are written by 
\begin{align}
    [g^0]_2[g^0]_2=[g^0]_2+[g^1]_2,\qquad [g^0]_2[g^1]_2=[g^0]_2+[g^1]_2, \qquad 
    [g^1]_2[g^1]_2=[g^0]_2.
\end{align}
These fusion rules are different from those with $m=0$, that is, the Fibonacci fusion rules.
This is equivalent to the fusion rules of S2-b  for CICY with $h^{1,1}=2$.

\newpage

\section{CICYs with $h^{1,1}=4$}
\label{app:h11=4}

Table \ref{CICY-h11=4} shows concrete CICYs with $h^{1,1}=4$ as CICY-\# following \cite{CICY}.

     \begin{longtable}[h]{|c||p{12cm}|}
      \caption{Types of CICYs with $h^{1,1}=4$.} \label{CICY-h11=4}\\
        \hline
        Type & \multicolumn{1}{c|}{Number of CICY}\\
        \hline
        \endhead
        \addtocounter{c3}{1}
        \arabic{c3} &
        $ 4742 ,$ 
        $ 6200 ,$ 
        $ 6681 ,$ 
        $ 7044 ,$ 
        $ 7318 ,$ 
        $ 7415 ,$ 
        $ 7572 ,$ 
        $ 7573 ,$ 
        $ 7785 $ 
        \\
        \hline
        \addtocounter{c3}{1}
        \arabic{c3} &
        $ 4756 ,$ 
        $ 5615 ,$ 
        $ 6228 ,$ 
        $ 6232 ,$ 
        $ 6560 ,$ 
        $ 6562 ,$ 
        $ 6832 ,$ 
        $ 6833 ,$ 
        $ 7076 $ 
        \\
        \hline
        \addtocounter{c3}{1}
        \arabic{c3} &
        $ 5245 ,$ 
        $ 5776 ,$ 
        $ 5784 ,$ 
        $ 5823 ,$ 
        $ 6328 ,$ 
        $ 6527 ,$ 
        $ 6556 ,$ 
        $ 6557 ,$ 
        $ 6632 ,$ 
        $ 6633 ,$ 
        $ 6634 ,$ 
        $ 6696 ,$ 
        $ 6698 ,$ 
        $ 6772 ,$ 
        $ 6778 ,$ 
        $ 6928 ,$ 
        $ 6929 ,$ 
        $ 6984 ,$ 
        $ 6989 ,$ 
        $ 6990 ,$ 
        $ 7070 ,$ 
        $ 7159 ,$ 
        $ 7163 ,$ 
        $ 7217 ,$ 
        $ 7221 ,$ 
        $ 7239 ,$ 
        $ 7248 ,$ 
        $ 7250 ,$ 
        $ 7276 ,$ 
        $ 7285 ,$ 
        $ 7291 ,$ 
        $ 7301 ,$ 
        $ 7348 ,$ 
        $ 7349 ,$ 
        $ 7352 ,$ 
        $ 7368 ,$ 
        $ 7373 ,$ 
        $ 7391 ,$ 
        $ 7400 ,$ 
        $ 7404 ,$ 
        $ 7425 ,$ 
        $ 7429 ,$ 
        $ 7431 ,$ 
        $ 7449 ,$ 
        $ 7476 ,$ 
        $ 7490 ,$ 
        $ 7495 ,$ 
        $ 7540 ,$ 
        $ 7541 ,$ 
        $ 7602 ,$ 
        $ 7616 ,$ 
        $ 7628 ,$ 
        $ 7630 ,$ 
        $ 7649 ,$ 
        $ 7744 ,$ 
        $ 7777 $ 
        \\
        \hline
        \addtocounter{c3}{1}
        \arabic{c3} &
        $ 5255 ,$ 
        $ 5795 ,$ 
        $ 5805 ,$ 
        $ 5806 ,$ 
        $ 6207 ,$ 
        $ 6545 ,$ 
        $ 6546 ,$ 
        $ 6548 ,$ 
        $ 6790 ,$ 
        $ 6798 ,$ 
        $ 6809 ,$ 
        $ 6969 ,$ 
        $ 6970 ,$ 
        $ 7047 ,$ 
        $ 7052 ,$ 
        $ 7057 ,$ 
        $ 7135 ,$ 
        $ 7137 ,$ 
        $ 7139 ,$ 
        $ 7140 ,$ 
        $ 7215 ,$ 
        $ 7224 ,$ 
        $ 7353 ,$ 
        $ 7358 ,$ 
        $ 7416 ,$ 
        $ 7531 ,$ 
        $ 7532 ,$ 
        $ 7568 ,$ 
        $ 7634 ,$ 
        $ 7640 ,$ 
        $ 7748 $ 
        \\
        \hline
        \addtocounter{c3}{1}
        \arabic{c3} &
        $ 5304 ,$ 
        $ 5305 ,$ 
        $ 6222 ,$ 
        $ 6223 ,$ 
        $ 7073 ,$ 
        $ 7247 ,$ 
        $ 7489 ,$ 
        $ 7590 $ 
        \\
        \hline
        \addtocounter{c3}{1}
        \arabic{c3} &
        $ 5606 ,$ 
        $ 7043 ,$ 
        $ 7046 ,$ 
        $ 7457 ,$ 
        $ 7611 $ 
        \\
        \hline
        \addtocounter{c3}{1}
        \arabic{c3} &
        $ 5789 ,$ 
        $ 6171 ,$ 
        $ 6177 ,$ 
        $ 6206 ,$ 
        $ 6224 ,$ 
        $ 6226 ,$ 
        $ 6402 ,$ 
        $ 6403 ,$ 
        $ 6674 ,$ 
        $ 6675 ,$ 
        $ 6791 ,$ 
        $ 6963 ,$ 
        $ 7035 ,$ 
        $ 7039 ,$ 
        $ 7041 ,$ 
        $ 7056 ,$ 
        $ 7059 ,$ 
        $ 7074 ,$ 
        $ 7075 ,$ 
        $ 7077 ,$ 
        $ 7134 ,$ 
        $ 7211 ,$ 
        $ 7214 ,$ 
        $ 7347 ,$ 
        $ 7359 ,$ 
        $ 7371 ,$ 
        $ 7564 $ 
        \\
        \hline
        \addtocounter{c3}{1}
        \arabic{c3} &
        $ 5794 ,$ 
        $ 6014 ,$ 
        $ 6529 ,$ 
        $ 6547 ,$ 
        $ 6677 ,$ 
        $ 6682 ,$ 
        $ 6797 ,$ 
        $ 6966 ,$ 
        $ 7054 ,$ 
        $ 7226 ,$ 
        $ 7319 ,$ 
        $ 7414 ,$ 
        $ 7417 ,$ 
        $ 7469 ,$ 
        $ 7667 ,$ 
        $ 7723 $ 
        \\
        \hline
        \addtocounter{c3}{1}
        \arabic{c3} &
        $ 5827 ,$ 
        $ 6023 ,$ 
        $ 6563 $ 
        \\
        \hline
        \addtocounter{c3}{1}
        \arabic{c3} &
        $ 6013 ,$ 
        $ 6524 ,$ 
        $ 6558 ,$ 
        $ 7040 ,$ 
        $ 7072 ,$ 
        $ 7200 ,$ 
        $ 7448 ,$ 
        $ 7685 $ 
        \\
        \hline
        \addtocounter{c3}{1}
        \arabic{c3} &
        $ 6172 ,$ 
        $ 6219 ,$ 
        $ 7034 ,$ 
        $ 7067 ,$ 
        $ 7068 ,$ 
        $ 7205 ,$ 
        $ 7245 ,$ 
        $ 7329 ,$ 
        $ 7432 ,$ 
        $ 7482 ,$ 
        $ 7557 ,$ 
        $ 7582 ,$ 
        $ 7684 $ 
        \\
        \hline
        \addtocounter{c3}{1}
        \arabic{c3} &
        $ 6179 ,$ 
        $ 6196 ,$ 
        $ 6532 ,$ 
        $ 6678 ,$ 
        $ 6799 ,$ 
        $ 6816 ,$ 
        $ 6835 ,$ 
        $ 6964 ,$ 
        $ 6965 ,$ 
        $ 6967 ,$ 
        $ 7138 ,$ 
        $ 7356 $ 
        \\
        \hline
        \addtocounter{c3}{1}
        \arabic{c3} &
        $ 6184 ,$ 
        $ 6539 ,$ 
        $ 6810 ,$ 
        $ 6811 ,$ 
        $ 6817 ,$ 
        $ 7061 ,$ 
        $ 7141 ,$ 
        $ 7142 ,$ 
        $ 7216 ,$ 
        $ 7320 ,$ 
        $ 7361 ,$ 
        $ 7418 $ 
        \\
        \hline
        \addtocounter{c3}{1}
        \arabic{c3} &
        $ 6274 ,$ 
        $ 6275 ,$ 
        $ 6309 ,$ 
        $ 6325 ,$ 
        $ 6434 ,$ 
        $ 6603 ,$ 
        $ 6776 ,$ 
        $ 6779 ,$ 
        $ 6831 ,$ 
        $ 6839 ,$ 
        $ 7049 ,$ 
        $ 7080 ,$ 
        $ 7109 ,$ 
        $ 7114 ,$ 
        $ 7160 ,$ 
        $ 7162 ,$ 
        $ 7280 ,$ 
        $ 7284 ,$ 
        $ 7286 ,$ 
        $ 7290 ,$ 
        $ 7325 ,$ 
        $ 7327 ,$ 
        $ 7354 ,$ 
        $ 7395 ,$ 
        $ 7402 ,$ 
        $ 7403 ,$ 
        $ 7426 ,$ 
        $ 7427 ,$ 
        $ 7428 ,$ 
        $ 7461 ,$ 
        $ 7468 ,$ 
        $ 7478 ,$ 
        $ 7492 ,$ 
        $ 7498 ,$ 
        $ 7505 ,$ 
        $ 7507 ,$ 
        $ 7525 ,$ 
        $ 7538 ,$ 
        $ 7574 ,$ 
        $ 7593 ,$ 
        $ 7606 ,$ 
        $ 7653 ,$ 
        $ 7739 ,$ 
        $ 7829 $ 
        \\
        \hline
        \addtocounter{c3}{1}
        \arabic{c3} &
        $ 6784 ,$ 
        $ 6828 ,$ 
        $ 6933 ,$ 
        $ 7038 ,$ 
        $ 7042 ,$ 
        $ 7110 ,$ 
        $ 7113 ,$ 
        $ 7168 ,$ 
        $ 7199 ,$ 
        $ 7204 ,$ 
        $ 7218 ,$ 
        $ 7241 ,$ 
        $ 7270 ,$ 
        $ 7310 ,$ 
        $ 7334 ,$ 
        $ 7346 ,$ 
        $ 7405 ,$ 
        $ 7407 ,$ 
        $ 7434 ,$ 
        $ 7435 ,$ 
        $ 7446 ,$ 
        $ 7451 ,$ 
        $ 7462 ,$ 
        $ 7491 ,$ 
        $ 7493 ,$ 
        $ 7516 ,$ 
        $ 7517 ,$ 
        $ 7522 ,$ 
        $ 7524 ,$ 
        $ 7528 ,$ 
        $ 7548 ,$ 
        $ 7605 ,$ 
        $ 7607 ,$ 
        $ 7618 ,$ 
        $ 7629 ,$ 
        $ 7632 ,$ 
        $ 7654 ,$ 
        $ 7663 ,$ 
        $ 7719 ,$ 
        $ 7722 ,$ 
        $ 7733 ,$ 
        $ 7736 ,$ 
        $ 7738 ,$ 
        $ 7742 ,$ 
        $ 7751 ,$ 
        $ 7754 ,$ 
        $ 7772 ,$ 
        $ 7813 ,$ 
        $ 7818 ,$ 
        $ 7825 ,$ 
        $ 7852 $ 
        \\
        \hline
        \addtocounter{c3}{1}
        \arabic{c3} &
        $ 6803 $ 
        \\
        \hline
        \addtocounter{c3}{1}
        \arabic{c3} &
        $ 6814 ,$ 
        $ 6815 ,$ 
        $ 6905 ,$ 
        $ 7048 ,$ 
        $ 7088 ,$ 
        $ 7089 ,$ 
        $ 7090 ,$ 
        $ 7104 ,$ 
        $ 7209 ,$ 
        $ 7210 ,$ 
        $ 7222 ,$ 
        $ 7223 ,$ 
        $ 7271 ,$ 
        $ 7274 ,$ 
        $ 7275 ,$ 
        $ 7281 ,$ 
        $ 7283 ,$ 
        $ 7292 ,$ 
        $ 7326 ,$ 
        $ 7355 ,$ 
        $ 7357 ,$ 
        $ 7389 ,$ 
        $ 7390 ,$ 
        $ 7394 ,$ 
        $ 7397 ,$ 
        $ 7399 ,$ 
        $ 7420 ,$ 
        $ 7421 ,$ 
        $ 7430 ,$ 
        $ 7458 ,$ 
        $ 7467 ,$ 
        $ 7477 ,$ 
        $ 7506 ,$ 
        $ 7515 ,$ 
        $ 7519 ,$ 
        $ 7520 ,$ 
        $ 7539 ,$ 
        $ 7569 ,$ 
        $ 7596 ,$ 
        $ 7601 ,$ 
        $ 7604 ,$ 
        $ 7614 ,$ 
        $ 7639 ,$ 
        $ 7670 ,$ 
        $ 7672 ,$ 
        $ 7674 ,$ 
        $ 7681 ,$ 
        $ 7690 ,$ 
        $ 7700 ,$ 
        $ 7715 ,$ 
        $ 7767 ,$ 
        $ 7786 ,$ 
        $ 7805 ,$ 
        $ 7814 ,$ 
        $ 7820 ,$ 
        $ 7849 $ 
        \\
        \hline
        \addtocounter{c3}{1}
        \arabic{c3} &
        $ 7045 ,$ 
        $ 7471 ,$ 
        $ 7472 ,$ 
        $ 7784 $ 
        \\
        \hline
        \addtocounter{c3}{1}
        \arabic{c3} &
        $ 7060 ,$ 
        $ 7227 ,$ 
        $ 7230 ,$ 
        $ 7360 ,$ 
        $ 7475 ,$ 
        $ 7533 ,$ 
        $ 7724 ,$ 
        $ 7757 $ 
        \\
        \hline
        \addtocounter{c3}{1}
        \arabic{c3} &
        $ 7277 ,$ 
        $ 7278 ,$ 
        $ 7392 ,$ 
        $ 7423 ,$ 
        $ 7424 ,$ 
        $ 7510 ,$ 
        $ 7521 ,$ 
        $ 7597 ,$ 
        $ 7599 ,$ 
        $ 7652 ,$ 
        $ 7656 ,$ 
        $ 7657 ,$ 
        $ 7671 ,$ 
        $ 7691 ,$ 
        $ 7693 ,$ 
        $ 7737 ,$ 
        $ 7768 ,$ 
        $ 7775 ,$ 
        $ 7811 ,$ 
        $ 7835 ,$ 
        $ 7850 ,$ 
        $ 7856 $ 
        \\
        \hline
        \addtocounter{c3}{1}
        \arabic{c3} &
        $ 7445 ,$ 
        $ 7483 ,$ 
        $ 7675 ,$ 
        $ 7790 $ 
        \\
        \hline
        \addtocounter{c3}{1}
        \arabic{c3} &
        $ 7453 ,$ 
        $ 7454 ,$ 
        $ 7456 ,$ 
        $ 7615 ,$ 
        $ 7749 ,$ 
        $ 7796 ,$ 
        $ 7837 $ 
        \\
        \hline
        \addtocounter{c3}{1}
        \arabic{c3} &
        $ 7455 ,$ 
        $ 7518 ,$ 
        $ 7682 $ 
        \\
        \hline
        \addtocounter{c3}{1}
        \arabic{c3} &
        $ 7470 ,$ 
        $ 7683 $ 
        \\
        \hline
        \addtocounter{c3}{1}
        \arabic{c3} &
        $ 7473 $ 
        \\
        \hline
        \addtocounter{c3}{1}
        \arabic{c3} &
        $ 7497 ,$ 
        $ 7598 ,$ 
        $ 7650 ,$ 
        $ 7692 ,$ 
        $ 7740 ,$ 
        $ 7793 ,$ 
        $ 7827 $ 
        \\
        \hline
        \addtocounter{c3}{1}
        \arabic{c3} &
        $ 7839 $ 
        \\
        \hline
        \addtocounter{c3}{1}
        \arabic{c3} &
        $ 7857 $ 
        \\
        \hline
        \addtocounter{c3}{1}
        \arabic{c3} &
        $ 7859 $ 
        \\
        \hline
        \addtocounter{c3}{1}
        \arabic{c3} &
        $ 7860 $ 
        \\
        \hline
        \addtocounter{c3}{1}
        \arabic{c3} &
        $ 7862 $ 
        \\
        \hline
        \addtocounter{c3}{1}
        \arabic{c3} &
        $ 7864 $ 
        \\
        \hline
    \end{longtable}

\newpage

Tables \ref{tab:symmetry-h11=4} and \ref{tab:type28-h11=4} show  symmetries for CICYs with $h^{1,1}=4$.
Wee need a new type of fusion rules, S4 in Eq.~(\ref{eq:S4}) to understand coupling selection rules of Types 20 and 28.
That is the reason why the asterisk symbol is put for Types 20 and 28 in Tables \ref{tab:symmetry-h11=4} and \ref{tab:type28-h11=4}. 
Selection rules of types shown in Table \ref{tab:symmetry-h11=4} can be understood by four combinations of S1, S2-a, S2-b, S3, S4, while we need their five combinations for Type 28 as shown in Table \ref{tab:type28-h11=4}.

{\tiny
\begin{longtable}[H]{|c||c|c|c|c|c|}
\caption{Symmetries for $h^{1,1}=4$.}
\label{tab:symmetry-h11=4}\\
    \hline
    Type & Symmetries & $A_1$ & $A_2$ & $A_3$ & $A_4$\\
    \hline
1	&(S2-b)$\times$(S3)$\times$(S2-a)$\times$(S2-a)	&$[b_{2,1}][a_{3,1}][a_{2,2}][a_{2,2}]$	&$[b_{2,2}][a_{3,2}][a_{2,2}][a_{2,2}]$	&$[b_{2,2}][a_{3,3}][a_{2,1}][a_{2,2}]$	&$[b_{2,2}][a_{3,3}][a_{2,2}][a_{2,1}]$	\\	\hline
2	&(S2-a)$\times$(S2-a)$\times$(S2-a)$\times$(S1)	&$[a_{2,1}][a_{2,2}][a_{2,2}][a_{1}]$	&$[a_{2,2}][a_{2,1}][a_{2,2}][a_{1}]$	&$[a_{2,2}][a_{2,2}][a_{2,1}][a_{1}]$	&$[a_{2,2}][a_{2,2}][a_{2,2}][a_{1}]$	\\	\hline
3	&(S2-b)$\times$(S2-b)$\times$(S2-a)$\times$(S1)	&$[b_{2,1}][b_{2,2}][a_{2,2}][a_{1}]$	&$[b_{2,2}][b_{2,1}][a_{2,2}][a_{1}]$	&$[b_{2,2}][b_{2,2}][a_{2,1}][a_{1}]$	&$[b_{2,2}][b_{2,2}][a_{2,2}][a_{1}]$	\\	\hline
4	&(S2-b)$\times$(S3)$\times$(S2-a)$\times$(S1)	&$[b_{2,1}][a_{3,1}][a_{2,2}][a_{1}]$	&$[b_{2,2}][a_{3,2}][a_{2,2}][a_{1}]$	&$[b_{2,2}][a_{3,3}][a_{2,1}][a_{1}]$	&$[b_{2,2}][a_{3,3}][a_{2,2}][a_{1}]$	\\	\hline
5	&(S2-a)$\times$(S2-a)$\times$(S2-a)$\times$(S2-a)	&$[a_{2,1}][a_{2,2}][a_{2,2}][a_{2,2}]$	&$[a_{2,2}][a_{2,1}][a_{2,2}][a_{2,2}]$	&$[a_{2,2}][a_{2,2}][a_{2,1}][a_{2,2}]$	&$[a_{2,2}][a_{2,2}][a_{2,2}][a_{2,1}]$	\\	\hline
6	&(S2-b)$\times$(S2-a)$\times$(S3)$\times$(S2-a)	&$[b_{2,1}][a_{2,2}][a_{3,1}][a_{2,2}]$	&$[b_{2,2}][a_{2,1}][a_{3,3}][a_{2,2}]$	&$[b_{2,2}][a_{2,2}][a_{3,2}][a_{2,2}]$	&$[b_{2,2}][a_{2,2}][a_{3,3}][a_{2,1}]$	\\
\hline
\end{longtable}}

\begin{table}[H]
    \centering
    \caption{Symmetries in Type 28 for $h^{1,1}=4$.}
        \label{tab:type28-h11=4}
    \begin{tabular}{|c|c|}
    \hline
        Type & $28^*$ \\
        \hline
        Symmetries & (S2-a)$\times$(S3)$\times$ (S3)$\times$ (S3)$\times$ (S4)\\
        \hline
        $A_1$ & $[a_{2,2}][a_{3,2}][a_{3,1}][a_{3,1}][a_{3,1}]$	\\
        \hline
        $A_2$ & $[a_{2,2}][a_{3,1}][a_{3,2}][a_{3,1}][a_{3,1}]$	\\
        \hline
        $A_3$ & $[a_{2,2}][a_{3,3}][a_{3,3}][a_{3,2}][a_{3,2}]$	\\
        \hline
        $A_4$ & $[a_{2,1}][a_{3,1}][a_{3,1}][a_{3,3}][a_{3,3}]$	\\
        \hline
    \end{tabular}
\end{table}

\newpage

    Table \ref{texture-h11=4}  shows possible textures for CICYs 
 with $h^{1,1}=4$ when the Higgs field $H$ corresponds to $A_1$, $A_2$, $A_3$, or $A_4$.

  \begin{longtable}[h]{|c||c|c|c|c|}
  \caption{Yukawa textures from CICYs with $h^{1,1}=4$.} \label{texture-h11=4}\\
            \hline
            Type \textbackslash $H$ & $A_1$ & $A_2$ & $A_3$ & $A_4$\\
            \hline
            \endhead
            
                        \addtocounter{c1}{1}
            Type \arabic{c1} 
            & $\pmqty{
                 0  & 
                 0  & 
                 0  & 
                 0 \\ 
                 0  & 
                 0  & 
                 *  & 
                 * \\ 
                 0  & 
                 *  & 
                 *  & 
                 * \\ 
                 0  & 
                 *  & 
                 *  & 
                 * \\ 
            }$
            & $\pmqty{
                 0  & 
                 0  & 
                 *  & 
                 * \\ 
                 0  & 
                 0  & 
                 *  & 
                 * \\ 
                 *  & 
                 *  & 
                 *  & 
                 * \\ 
                 *  & 
                 *  & 
                 *  & 
                 * \\ 
            }$
            & $\pmqty{
                 0  & 
                 *  & 
                 *  & 
                 * \\ 
                 *  & 
                 *  & 
                 *  & 
                 * \\ 
                 *  & 
                 *  & 
                 0  & 
                 * \\ 
                 *  & 
                 *  & 
                 *  & 
                 * \\ 
            }$
            & $\pmqty{
                 0  & 
                 *  & 
                 *  & 
                 * \\ 
                 *  & 
                 *  & 
                 *  & 
                 * \\ 
                 *  & 
                 *  & 
                 *  & 
                 * \\ 
                 *  & 
                 *  & 
                 *  & 
                 0 \\ 
            }$
            \\
\hline
     \end{longtable}

\newpage

\section{CICYs with $h^{1,1}=5$}
\label{app:h11=5}

Table \ref{tab:prepotential-h11=5} shows classification of coupling selection rules for CICYs with $h^{1,1}=5$.

\begin{longtable}[H]{|c||p{12cm}|}
\caption{Types of prepotential for CICYs with $h^{1,1}=5$. Vanishing $\kappa_{abc}$ are shown.} \label{tab:prepotential-h11=5} \\
\hline
Type & \multicolumn{1}{c|}{$\kappa_{abc}=0$} \\
\hline
\endhead
1 & 
$\kappa_{ 1 1 1 }$,
$\kappa_{ 1 1 2 }$,
$\kappa_{ 1 1 3 }$,
$\kappa_{ 1 1 4 }$,
$\kappa_{ 1 1 5 }$,
$\kappa_{ 1 2 2 }$,
$\kappa_{ 1 3 3 }$,
$\kappa_{ 2 2 2 }$,
$\kappa_{ 2 2 3 }$,
$\kappa_{ 2 2 4 }$,
$\kappa_{ 2 2 5 }$,
$\kappa_{ 2 3 3 }$,
$\kappa_{ 3 3 3 }$,
$\kappa_{ 3 3 4 }$,
$\kappa_{ 3 3 5 }$
\\ \hline
2 & 
$\kappa_{ 1 1 1 }$,
$\kappa_{ 1 1 2 }$,
$\kappa_{ 1 1 3 }$,
$\kappa_{ 1 1 4 }$,
$\kappa_{ 1 1 5 }$,
$\kappa_{ 1 2 2 }$,
$\kappa_{ 1 3 3 }$,
$\kappa_{ 2 2 2 }$,
$\kappa_{ 2 2 3 }$,
$\kappa_{ 2 2 4 }$,
$\kappa_{ 2 2 5 }$,
$\kappa_{ 2 3 3 }$,
$\kappa_{ 3 3 3 }$,
$\kappa_{ 3 3 4 }$,
$\kappa_{ 3 3 5 }$,
$\kappa_{ 4 4 4 }$,
$\kappa_{ 5 5 5 }$
\\ \hline
3 & 
$\kappa_{ 1 1 1 }$,
$\kappa_{ 1 1 2 }$,
$\kappa_{ 1 1 3 }$,
$\kappa_{ 1 1 4 }$,
$\kappa_{ 1 1 5 }$,
$\kappa_{ 1 2 2 }$,
$\kappa_{ 1 3 3 }$,
$\kappa_{ 2 2 2 }$,
$\kappa_{ 2 2 3 }$,
$\kappa_{ 2 2 4 }$,
$\kappa_{ 2 2 5 }$,
$\kappa_{ 3 3 3 }$,
$\kappa_{ 4 4 4 }$
\\ \hline
4 & 
$\kappa_{ 1 1 1 }$,
$\kappa_{ 1 1 2 }$,
$\kappa_{ 1 1 3 }$,
$\kappa_{ 1 1 4 }$,
$\kappa_{ 1 1 5 }$,
$\kappa_{ 1 3 3 }$,
$\kappa_{ 1 4 4 }$,
$\kappa_{ 2 2 2 }$,
$\kappa_{ 3 3 3 }$,
$\kappa_{ 4 4 4 }$,
$\kappa_{ 5 5 5 }$
\\ \hline
5 & 
$\kappa_{ 1 1 1 }$,
$\kappa_{ 1 1 2 }$,
$\kappa_{ 1 1 3 }$,
$\kappa_{ 1 1 4 }$,
$\kappa_{ 1 1 5 }$,
$\kappa_{ 1 4 4 }$,
$\kappa_{ 2 2 2 }$,
$\kappa_{ 3 3 3 }$,
$\kappa_{ 4 4 4 }$
\\ \hline
6 & 
$\kappa_{ 1 1 1 }$,
$\kappa_{ 1 1 2 }$,
$\kappa_{ 1 1 3 }$,
$\kappa_{ 1 1 4 }$,
$\kappa_{ 1 1 5 }$,
$\kappa_{ 1 3 3 }$,
$\kappa_{ 2 2 2 }$,
$\kappa_{ 3 3 3 }$,
$\kappa_{ 4 4 4 }$
\\ \hline
7 & 
$\kappa_{ 1 1 1 }$,
$\kappa_{ 1 1 2 }$,
$\kappa_{ 1 1 3 }$,
$\kappa_{ 1 1 4 }$,
$\kappa_{ 1 1 5 }$,
$\kappa_{ 1 2 2 }$,
$\kappa_{ 1 4 4 }$,
$\kappa_{ 2 2 2 }$,
$\kappa_{ 3 3 3 }$,
$\kappa_{ 4 4 4 }$
\\ \hline
8 & 
$\kappa_{ 1 1 1 }$,
$\kappa_{ 1 1 2 }$,
$\kappa_{ 1 1 3 }$,
$\kappa_{ 1 1 4 }$,
$\kappa_{ 1 1 5 }$,
$\kappa_{ 1 2 2 }$,
$\kappa_{ 2 2 2 }$,
$\kappa_{ 3 3 3 }$,
$\kappa_{ 4 4 4 }$
\\
\hline

\end{longtable}

\newpage

Table \ref{CICY-h11=5} shows concrete CICYs with $h^{1,1}=5$ as CICY-\# following \cite{CICY}.

     \begin{longtable}[h]{|c||p{12cm}|}
        \caption{Types of concrete CICYs with $h^{1,1}=5$.} \label{CICY-h11=5}\\
        \hline
        Type & \multicolumn{1}{c|}{Number of CICY}\\
        \hline
        \endhead
        \addtocounter{c4}{1}
        \arabic{c4} &
        $ 3901 ,$ 
        $ 4554 ,$ 
        $ 5136 ,$ 
        $ 5137 ,$ 
        $ 5138 ,$ 
        $ 5139 ,$ 
        $ 5152 ,$ 
        $ 5564 ,$ 
        $ 5758 ,$ 
        $ 5775 ,$ 
        $ 5896 ,$ 
        $ 5914 ,$ 
        $ 5966 ,$ 
        $ 6078 ,$ 
        $ 6079 ,$ 
        $ 6092 ,$ 
        $ 6146 ,$ 
        $ 6204 ,$ 
        $ 6278 ,$ 
        $ 6280 ,$ 
        $ 6379 ,$ 
        $ 6457 ,$ 
        $ 6458 ,$ 
        $ 6468 ,$ 
        $ 6469 ,$ 
        $ 6479 ,$ 
        $ 6716 ,$ 
        $ 6717 ,$ 
        $ 6724 ,$ 
        $ 6727 ,$ 
        $ 6739 ,$ 
        $ 6744 ,$ 
        $ 6746 ,$ 
        $ 6758 ,$ 
        $ 6768 ,$ 
        $ 6789 ,$ 
        $ 6856 ,$ 
        $ 6914 ,$ 
        $ 6926 ,$ 
        $ 7005 ,$ 
        $ 7006 ,$ 
        $ 7007 ,$ 
        $ 7029 ,$ 
        $ 7107 ,$ 
        $ 7174 ,$ 
        $ 7181 ,$ 
        $ 7198 ,$ 
        $ 7279 ,$ 
        $ 7289 ,$ 
        $ 7339 ,$ 
        $ 7408 ,$ 
        $ 7621 $ 
        \\
        \hline
        \addtocounter{c4}{1}
        \arabic{c4} &
        $ 4064 ,$ 
        $ 4862 ,$ 
        $ 5684 ,$ 
        $ 5783 ,$ 
        $ 6175 ,$ 
        $ 6176 ,$ 
        $ 6225 ,$ 
        $ 6526 $ 
        \\
\hline
    \end{longtable}

Coupling selection rules in most of types can be understood by 
five products of S1, S2-a, S2-b, S3, and S4. However, we need six products of S1, S2-a, S2-b, S3, and S4 for 
Types 11, 53, 65, 68, 70, 73, 75, 79, 80, 82, 85, 86, 89, 90, 91, and 92, and seven products for Types 87 and 88.

\bibliography{references}{}
\bibliographystyle{JHEP}

\end{document}